\documentclass[authorversion,nonacm]{acmart}

\AtBeginDocument{%
  }

\usepackage{bm}
\usepackage{color}
\usepackage{booktabs}
\usepackage{multirow}
\usepackage{makecell}

\usepackage{amsmath}
\DeclareMathOperator*{\argmax}{arg\,max}

\DeclareMathOperator*{\E}{\mathbb{E}}

\usepackage{enumitem}

\begin{document}

\title{Co-Training Realized Volatility Prediction Model with Neural Distributional Transformation}
\thanks{Accepted at ICAIF'23}

\author{Xin Du}
\orcid{1234-5678-9012}
\authornotemark[1]
\affiliation{%
  \institution{Waseda University}
  \streetaddress{3-chōme-4-1 Ōkubo, Shinjuku City}
  \city{Tokyo}
  \country{Japan}
  \postcode{169-0072}
}
\email{duxin@aoni.waseda.jp}

\author{Kai Moriyama}
\orcid{0009-0007-0530-2082}
\affiliation{%
  \institution{Waseda University}
  \city{Tokyo}
  \country{Japan}}
\email{scotify@akane.waseda.jp}

\author{Kumiko Tanaka-Ishii}
\orcid{0000-0003-1752-3951}
\affiliation{%
  \institution{Waseda University}
  \city{Tokyo}
  \country{Japan}
}
\email{kumiko@waseda.jp}

\begin{abstract}

This paper shows a novel machine learning model for realized
volatility (RV) prediction using a {\em normalizing flow}, an
invertible neural network. Since RV is known to be skewed and have a
fat tail, previous methods transform RV into values that follow a
latent distribution with an explicit shape and then apply a prediction
model. However, knowing that shape is non-trivial, and the
transformation result influences the prediction model. This paper
proposes to jointly train the transformation and the prediction model.
The training process follows a maximum-likelihood objective function
that is derived from the assumption that the prediction residuals on
the transformed RV time series are homogeneously Gaussian. The
objective function is further approximated using an
expectation–maximum algorithm. On a dataset of 100 stocks, our method
significantly outperforms other methods using analytical or naïve
neural-network transformations.

\end{abstract}

\ccsdesc[500]{Applied computing~Economics}
\ccsdesc[500]{Computing methodologies~Neural networks}
\ccsdesc[500]{Mathematics of computing~Time series analysis}

\keywords{realized volatility, neural networks, time-series prediction, normalizing flow}


\maketitle

\section{Introduction}

Volatility is the primary measure of financial risk, and its time
series modeling is an important task in financial engineering.
Volatility presents a large skew toward the tail, usually called ``fat
tail.'' Since modeling this distribution is non-trivial, previous
methods to analytically transform the distribution into a more
tractable latent distribution often presume a certain prior rigorous
distributional shape, such as Gaussian.

Previously, \citet{ghaddar1981data} proposed to {\em Gaussianize} the
distribution by a Box-Cox power transformation
\citep{box1964analysis}.  \citet{proietti2013does} showed, however,
that Box-Cox transformation outperformed the other transformations in
only a fraction of all time series studied.  Other analytical families
of transformations have been considered \citep{yeo2000new,
goerg2015lambert, taylor2017realised}, but choosing the ``correct''
shape is non-trivial, and the latent distribution might not have any
tractable shape.  Furthermore, this transformation is influenced by
what comes after the transformation (i.e., the prediction model).

Such questions naturally lead to an idea to transform volatility into
a {\em tuned} latent distribution that best fits the prediction model.
In other words, we are interested in {\em jointly} fitting the
transformation and the prediction model together. If the
volatility is explicit, then the joint training becomes tractable via
the residual of the prediction.

Therefore, we propose a new machine-learning model for volatility
prediction. We use realized volatility (RV
\citep{clark1973subordinated}) defined with high-frequency data as the
root sum of the quadratic variations of intraday high-frequency
price returns. While other definitions of volatility, including
conditional volatility \citep{engle1982autoregressive} or stochastic
volatility \citep{taylor1982financial}, remain implicit and require
statistical inference, RV has acquired popularity with an increasing
availability of high-frequency data.

\begin{figure*}[htbp]
  \centering
  \includegraphics[width=0.75\linewidth]{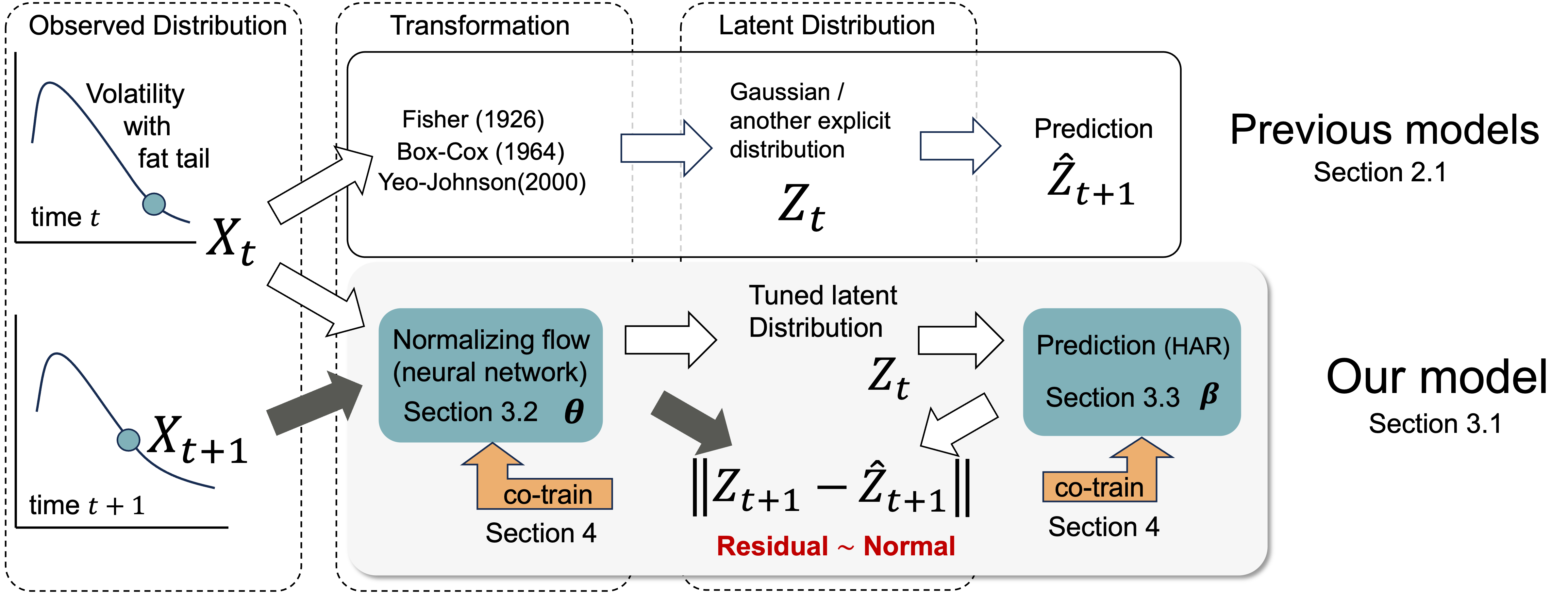}
  \caption{
  An overview of our approach (lower half) compared with
  the previous approach (upper half). Unlike previous approaches,
  our approach does not assume the
  latent distribution to be Gaussian or another tractable shape.
  Furthermore, the prediction model is co-trained 
  with the neural transformation.
  }
  \label{fig:overview}
\end{figure*}

Our approach is demonstrated by the lower half of Figure
\ref{fig:overview}. A volatility value $X_t$ at the timestep $t$ is
transformed with a neural network, specifically, a technique of
normalizing flow \citep{rezende2015variational}. Normalizing flow
realizes an invertible transformation, and it can be trained to acquire
the best approximation to a desired distribution.  The transformed
latent values $Z_{1},\cdots,Z_{t}$ are fed into a prediction model
that outputs the estimate at the following timestep $\hat{Z}_{t+1}$,
for which we use a {\em heterogeneous auto-regressive} (HAR) model
\citep{corsi2009har}. At the same time, we have the actual realized
volatility $X_{t+1}$ of time $t+1$, which is transformed into
$Z_{t+1}$. The residual $\hat{Z}_{t+1}-Z_{t+1}$ can be used to jointly
train both the transformation and prediction, naturally realized as an
expectation–maximization procedure.

We show how our setting outperforms other possibilities when using no
transformation or using analytical or simple neural-network
transformations.

\section{Related Works}
\label{sec:relatedworks}

First proposed in the 1970s \citep{clark1973subordinated}, with a
greater availability of high-frequency data, RV now
holds the key to many financial engineering tasks, such as the
prediction of return variation \citep{andersen2000exchange,
thomakos2003realized} and option pricing \citep{corsi2013realizing}.

This paper focuses on the problem of RV prediction with a
distributional transformation. Various analytical functions have been
considered previously, which we summarize in Section
\ref{sec:rvpred-transform}. While the transformation operates for each
timestep, it requires a time-series model to address the temporal
dependence across timesteps. Section \ref{sec:rvpred} introduces
several such time-series models. Moreover, our approach considers the
transformed RV as latent variables; Section \ref{sec:latent} compares
our approach with existing latent-variable models.

\subsection{Realized Volatility Prediction with Analytical
Transformations} \label{sec:rvpred-transform}

The use of distributional transformations dates back to
\citet{fisher1926expansion}, where a Student's $t$ variable was
expanded as an infinite series of normal variables. In another early
work, \citet{wallace1959bounds} proposed a simple approximation form
that was found to work exceptionally well transforming between
Gaussian and Student's $t$ distributions
\citep{prescott1974normalizing}.

A transformation can be used to improve the Gaussianity of a dataset
that significantly deviates from a Gaussian distribution. This is
often beneficial as Gaussianity occurs as an assumption underlying
many statistical and econometric models.

The modern research on distributional transformation for economic data
is largely based on \citet{box1964analysis}’s seminal work, which
introduced a family of power transformations now called Box-Cox
transformations. Box-Cox transformations include the identity
transformation (i.e., $f(x)=x$) and the log transformation as two
special cases. Box-Cox transformations are applied widely, but they
only allow positive inputs. Yeo-Johnson transformations
\citep{yeo2000new} were later introduced to generalize Box-Cox to
include zero and negative inputs. Apart from power transformations,
numerous alternatives have been proposed.
\citet{hoyle1973transformations} summarized 18 different analytical
transformations. Other choices include the Tukey $g$-and-$h$
\citep{tukey1977modern}, the Lambert W function
\citep{goerg2015lambert}, and exponential functions
\citep{yan2018parsimonious}.

While the improvement via a Gaussianizing transformation is frequently
observed for cross-sectional data (i.e., a collection of samples at a
given time), it does not apply to sequential data.
\citet{proietti2013does} tested Box-Cox transformations on various
real datasets with other transformations and found that Box-Cox
transformations showed advantages only in a fifth of all cases.
\citet{taylor2017realised} investigated the use of Box-Cox
transformations for RV prediction. Among several pre-defined choices
of the parameter $\lambda$ of the Box-Cox transformation,
$\lambda=1/4$ performed the best. However, as mentioned in the
Introduction, the ``correct'' shape of the transformed distribution is
influenced by the time-series prediction model. The optimal parameter
in one scenario might be sub-optimal in other scenarios.

This study uses neural networks for the transformation. Simple
neural-network transformations have been considered previously for
time-series modeling. \citet{snelson2003warped} proposed using an
additive mixture of tanh functions, a two-layer feed-forward neural
network. The parameters in the neural network were restricted to be
positive, so that the neural network remains invertible. Recent
state-of-the-art techniques are with normalizing flows
\citep{tabak2010density, dinh2015nice, rezende2015variational}, which
provide more powerful choices than the mixture-of-tanh approach in
\citet{snelson2003warped}.

\subsection{Realized Volatility Prediction Models}
\label{sec:rvpred}

Realized volatility, like other volatility measures, exhibits
long-memory characteristics often referred to as ``volatility
clustering'' \citep{corsi2008volatility}. Based on this finding, many
models have been proposed to predict RV, among which
two are especially popular. The first is the {\em heterogeneous
auto-regressive} (HAR) model by \citet{corsi2009har}, which uses a
linear combination of past volatility terms defined over different
time periods to predict future volatility. HAR is simple but
performs well in practice; therefore, we use it as the
prediction model in this paper. The other prevalent model is the {\em
auto-regressive fractionally integrated moving average} (ARFIMA)
\citep{granger1980introduction}, which utilizes fractional
differencing operators to enable long memory.

It must be noted, however, that RV prediction is closely related to
the fat tail problem in financial returns. Research on the fat tails
in financial returns dates to \citet{Mandelbrot1963TheVO} and
\citet{fama1965behavior}, who noticed that substantial price changes
occur more frequently than a Gaussian predicts; thus, the observed
return distribution shows a larger kurtosis than a Gaussian (i.e., it
has a fat tail). This observation was further validated on large-scale
measurements of high-frequency price data
\citep{gopikrishnan1998inverse, gopikrishnan1999scaling,
plerou1999scaling, gabaix2003theory}. Volatility prediction models
have been developed to account for the fat-tailed distribution of
returns. But as we will show via baselines in Section
\ref{sec:result}, this implicit approach via volatility models is
often insufficient to fully capture the fat tail.

\subsection{Latent Sequential Models}
\label{sec:latent}

The modeling of a transformed time series can be seen as a kind of
latent-variable model. Realized volatilities are ``observable''
variables while the transformed values are seen as ``latent.'' Because
of the long-memory nature of price time-series, the long-range
dependence between the latent variables must be addressed. How to
integrate financial econometrics methods into the latent-variable
framework of machine learning remains a question.

Several latent-variable methods are considered relevant to our
approach. The first is the Kalman filter \citep{kalman1961new}, which
assumes that all latent variables are Gaussian and governed by a
linear Markovian process. A Kalman filter is estimated by making an
initial guess of the earliest latent variable and then progressively
gets refined by incorporating new observations at later timesteps.

The second is the hidden Markov models (HMMs)
\citep{baum1966statistical}. HMMs are more elaborate latent-variable
models that have been used widely in various fields
\citep{cappe2005inference}. Kalman filters can be viewed as special
cases of HMMs, even though they were studied in different research
streams for a long time \citep{cappe2005inference}.

Nevertheless, the Markovian nature of Kalman filters and HMMs makes it
hard to capture the long memory in RV time-series data
\citep{corsi2009har}. In contrast, a simpler HAR can easily access a
timestep far back (e.g., 22 timesteps previous to current), which is
much more straightforward than HMMs in incorporating the long memory
and has been found to work well.

In this paper, we use normalizing flows (NFs) to assist HAR models in
capturing the complex RV distribution. NFs were proposed for
variational inference \citep{rezende2015variational}, which is a
method frequently used for latent-variable modeling. Thus far,
normalizing flows have been shown to be effective in various
applications, such as density estimation \citep{tabak2010density},
generative modeling \citep{dinh2015nice}, variational inference
\citep{rezende2015variational}, noise modeling
\citep{abdelhamed2019noise}, and time-series modeling
\citep{deng2020modeling}.

In a recent work by \citet{deng2020modeling}, normalizing flows were
used to transform a time series into a Brownian motion. As Brownian
motions are Markov processes, the model by \citet{deng2020modeling}
can be seen as a hidden Markov model in which the latent Gaussian
variables are estimated using a normalizing flow. In contrast, the
latent process in this work must capture the long-memory phenomena
observed in real RV data \citep{corsi2009har}.

\section{Realized Volatility Prediction via Transformation}

Daily RV is the root sum of quadratic price variations during a
trading day. RV measures the amount of financial risk during the day
and is calculated using high-frequency price records. Let $p_{t,i}$
denote the $i$-th high-frequency price of a stock of day $t$
($t=1,\cdots,T$). Then, the daily RV, $x_t$, is calculated as follows:
\begin{equation}
  x_t = \sqrt{\sum_{i=2}^{n_t} (\log p_{t,i} - \log p_{t,i-1})^2},
  \label{eq:rv}
\end{equation} where $n_t$ denotes the number of high-frequency prices
within day $t$.

The prediction of daily RV is to estimate $x_{t+1}$ using the previous
values $x_{\leq t}\equiv [x_1,\cdots,x_t]$. Let $y_{\bm{\beta}}$
denote a RV prediction model with parameters $\bm{\beta}$. The aim is
to find the best parameters that minimize the {\em root-mean-square
error} (RMSE) between the prediction and the true values, as follows:
\begin{equation}
  \min_{\bm{\beta}} \text{RMSE}(\bm{\beta})
  = \sqrt{\frac{1}{T-1}\sum_{t=1}^{T-1} (\hat{x}_{t+1} -
  x_{t+1})^2},
  \label{eq:rmse}
\end{equation}
where $\hat{x}_{t+1}=y_{\bm{\beta}}(x_{\leq t})$ denotes the predicted
value.

\subsection{Prediction with Transformed Time Series}
\label{sec:problem}

In this paper, we study the task of predicting RVs with a learnable
monotonic transformation $f_{\bm{\theta}}:\mathbb{R}\to\mathbb{R}$
parameterized by $\bm{\theta}$. At every timestep $t$, realized
volatility $x_t$ is transformed into $z_t=f_{\bm{\theta}}(x_t)$, which
follows a latent distribution. Instead of predicting $x_{t+1}$
directly, $z_{t+1}$ is predicted as $\hat{z}_{t+1}$ using $z_{\leq
t}$, and the realized volatility $x_{t+1}$ is then estimated from
$\hat{z}_{t+1}$. The entire prediction procedure is as follows:
\begin{align}
  \hat{x}_{t+1} &= \mathbb{E}[f^{-1}_{\bm{\theta}}(\hat{z}_{t+1})],
  \label{eq:pred-inverse} \\
  \hat{z}_{t+1} &= y_{\bm{\beta}} (z_{\leq t}),
  \label{eq:pred-autoregress}\\
  z_s &= f_{\bm{\theta}}(x_s) \qquad s=1,2,\cdots,t,
  \label{eq:pred-transform}
\end{align} where $y_{\bm{\beta}}$ denotes the RV prediction model;
$\mathbb{E}$ means to take the average when $\hat{z}_{t+1}$ is a
random variable. Here, we assume that $\hat{z}_{t+1}$ is
non-stochastic and $\mathbb{E}[f^{-1}(\hat{z}_{t+1})]$ reduces to
$f^{-1}(\hat{z}_{t+1})$.

Using the transformation $f_{\bm{\theta}}$ has multiple advantages.
First, the transformed realized volatilities better approximate
Gaussian distributions, an implicit assumption that underlies many
prediction models. Thus, the prediction model is estimated with less
bias. Second, for a linear prediction model $y_{\bm{\beta}}$, the use
of nonlinear transformation $f_{\bm{\theta}}$ enables capturing
nonlinear temporal dependency.

\subsection{Normalizing Flow}
\label{sec:node}

In this paper, we propose implementing the invertible transformation
$f_{\bm{\theta}}$ with {\em normalizing flow}, a deep-learning
technique. Normalizing flow refers to invertible neural networks for
which the inverse can be efficiently computed. Specifically, we use
{\em neural ordinary differential equations} (NODE)
\citep{chen2018neural}, a special kind of normalizing flow.

A NODE is a first-order ordinary differential equation parameterized
by a neural network:
\begin{equation}
  \frac{\mathrm{d}g(x,\xi)}{\mathrm{d}\xi}
  = \text{NN}_{\bm{\theta}}(g(x,\xi), \xi).
  \label{eq:node}
\end{equation}
When used as a normalizing flow, the input is set to $g(x, 0)$, and
the value of $g(x, \tau)$ ($\tau>0$) acquired as follows is used as
the output. For $t=1,\cdots,T$,
\begin{align}
  x_t &= g(x_t, 0) \\
  z_t &= f_{\bm{\theta}}(x_t) = g(x_t, \tau),
\end{align} where $\tau$ is a hyperparameter that controls the
complexity of the transformation. The existence and uniqueness of
$g(x,\tau)$ given $g(x,0)$ is guaranteed by the Picard-Lindelöf
theorem, as is $g(x,0)$ given $g(x,\tau)$ for the case of computing
$f_{\bm{\theta}}^{-1}$. \citet{chen2018neural} proposed efficient
numerical methods to calculate the gradient $\nabla_{\bm{\theta}}
f_{\bm{\theta}}$ (or $f^\prime _{\bm{\theta}}$ in our case with 1D
transformations), which is used for gradient-based optimization of
$\bm{\theta}$.

\subsection{Heterogeneous Auto-Regression}

For the prediction model $y_{\bm{\beta}}$, we use the {\em
  heterogeneous auto-regressive} (HAR) model \citep{corsi2009har}. HAR
is one of the most common models for RV prediction.

HAR is a linear regression model that incorporates RV components
during three different periods, as follows:
\begin{equation}
  z_{t+1} = \beta_0 + \beta^\text{(d)} z_t^\text{(d)} +
  \beta^\text{(w)} z_t^\text{(w)} + \beta^{(m)} z_t^\text{(m)} +
  \varepsilon_{t+1},
  \label{eq:har}
\end{equation} where $z_t^\text{(d)}\equiv z_t$, $z_t^\text{(w)}\equiv
z_{t-4}+\cdots+z_{t}$, and $z_t^\text{(m)}\equiv z_{t-21}+\cdots+z_t$
represent the previous day, the previous week, and the previous month,
respectively. $\varepsilon_{t+1}$ is the residual. The parameters can
be estimated with an ordinary least-square estimator. When HAR is used
without a transformation or with an identity transformation $f(x) =
x$, the residuals $\varepsilon_{t+1}$ typically follow a skewed
distribution.

Although this paper focuses on HAR, the proposed method can be applied
to prediction models other than HAR. Because normalizing flow
specifies a wide family of continuous functions, it requires little
knowledge about the prediction model to effectuate the prediction in
Formulas (\ref{eq:pred-inverse}) through (\ref{eq:pred-transform}).
The universality of our method with different prediction models remain
a future work.

\section{Co-training Transformation with Prediction Model}

Previously, the estimation of transformation parameters was done prior
to that of the prediction model, following some sub-optimal
objectives, such as the presumed Gaussianity of the transformed data.

In the following, we propose a method to co-train (i.e., jointly
train) the transformation with the prediction model. A primary
challenge in this approach arises when the transformation and
prediction components employ different techniques, like neural
networks and linear regression in our case. Existing literature has
yet to present a unified framework that optimizes both components
together. This section details our solution to address this challenge.

\subsection{Parameter Estimation via Expectation Maximization}

Let $X=[X_1, X_2, \cdots]$ and $Z=[Z_1, Z_2, \cdots]$ denote the
random processes underlying the raw and the transformed realized
volatilities, respectively, where $Z_s=f_{\bm{\theta}}(X_s)$
($s=1,2,\cdots$). The normalizing flow $f_{\bm{\theta}}$ is optimized
subject to a maximum-likelihood objective as follows:
\begin{align}
  \max_{\bm{\theta}} \log\mathrm{P}(X\mid \bm{\theta})
  &= \log\int_Z \mathrm{P}(X, Z\mid \bm{\theta})~\mathrm{d}Z \\
  &= \log\int_Z \mathrm{P}(X\mid Z, \bm{\theta})~\mathrm{P}(Z\mid \bm{\theta})~\mathrm{d}Z \\
  &= \log\E_{Z\sim \mathrm{P}(\cdot\mid \bm{\theta})} \mathrm{P}(X\mid Z, \bm{\theta}),
  \label{eq:expectation}
\end{align}
which is estimated with the empirical average of $\mathrm{P}(X\mid Z,
\bm{\theta})$ with respect to sample sequences $Z$ drawn from
$\mathrm{P}(\cdot\mid \bm{\theta})$. However, without knowing the true
parameters $\bm{\theta}$, the calculation is intractable. A common way
to handle this problem is via the {\em expectation maximization} (EM)
algorithm \citep{dempster1977maximum}, which starts from a raw estimate
$\bm{\theta}^{(0)}$ and refines the estimate iteratively as follows:
\begin{equation}
  \bm{\theta}^{(i+1)} = \argmax_{\bm{\theta}}
  \E_{Z\sim \mathrm{P}(\cdot\mid X, \bm{\theta}^{(i)})}
  \log \mathrm{P}(X\mid Z, \bm{\theta}).
  \label{eq:em}
\end{equation} The right-hand-side of Formula (\ref{eq:em}) is a lower
bound of Formula (\ref{eq:expectation}) because of the concavity of
the log function and Jensen's inequality. In other words, The
maximization of the log-likelihood in Formula (\ref{eq:expectation})
is effectuated through a maximization of an lower bound of the
log-likelihood as presented in Formula (\ref{eq:em}).

In this work, the dependence of $Z_t$ on $X_t$ is deterministic via
the normalizing flow, while that of $X_{t+1}$ on $Z_t$ can be random.
Hence, the sampling procedure of $Z\sim\mathrm{P}(\cdot\mid X,
\bm{\theta}^{(i)})$ in Formula (\ref{eq:em}) is simply to apply
$f_{\bm{\theta}}$ to every timestep of $X_1, X_2, \cdots$, and the
expectation average is taken over the RV time series of multiple
stocks. On the other hand, the term within the expectation operation
of Formula (\ref{eq:em}) decomposes into the following due to the
change-of-variable theorem:
\begin{align}
  \log\mathrm{P}(X\mid Z,\bm{\theta})
  &=\sum_{s} \log\mathrm{P}(X_{s+1}\mid Z_{\leq s},\bm{\theta}) \\
  &=\sum_{s} \log\left[
    \mathrm{P}(Z_{s+1}\mid Z_{\leq s})
    \cdot \left|f_{\bm{\theta}}^\prime(X_{t+1})\right|
  \right], \label{eq:decompose}
\end{align}
where $f_{\bm{\theta}}^\prime$ represents the first derivative of the
transformation, which a normalizing flow typically provides at a small
cost.

Notice that in Formula (\ref{eq:decompose}), $\mathrm{P}(Z_{s+1}\mid
Z_{\leq s})$ represents the prediction model $y_{\bm{\beta}}$ applied
to the latent process, and it does not specify how the parameters
$\bm{\beta}$ should be estimated. Therefore, optimization of the
parameters in the prediction model (i.e., $\bm{\beta}$) can be
effectuated via a different objective function from that in Formula
(\ref{eq:em}). This is important because HAR and many other financial
time-series models are conventionally estimated with a least-square
objective function, which imposes a weaker assumption on the
distribution of residuals and often shows better robustness. The
probability density $\mathrm{P}(Z_{s+1}\mid Z_{\leq s})$ is obtained
via an estimation, which is detailed in Section \ref{sec:residual}.

Hence, the co-training procedure of $f_{\bm{\theta}}$ and
$y_{\bm{\beta}}$ is via iterative updates as follows: at each
iteration, the sampling step $Z\sim P(\cdot\mid X, \bm{\theta}^{(i)})$
in Formula (\ref{eq:em}) is conducted as a random sampling of $X_{\leq
t}$ from the RV time series of stocks and then transforming $X_{\leq
t}$ into $Z_{\leq t}$. Using the samples of $Z_{\leq t}$, the
parameters of the prediction model (i.e., $\beta$) can be estimated
via any adequate method; here for HAR, we used the Python package
\texttt{arch}\footnote{\url{https://github.com/bashtage/arch}} to
estimate $\bm{\beta}$. Then, $Z_{t+1}$ is predicted, and the
expectation of log-likelihoods $\log P(x_{t+1}\mid Z_{\leq t},
\theta)$ is calculated using Formula (\ref{eq:decompose}). Finally,
the $\argmax$ operation in Formula (\ref{eq:em}) is approximated by a
mini-batch gradient ascent step over the expectation of the
log-likelihoods.

\subsection{Residual Density Estimation}
\label{sec:residual}

The conditional probability density $\mathrm{P}(Z_{s+1}\mid Z_{\leq
s})$ is determined by the prediction model $y_{\bm{\beta}}$, which
produces an estimate $\hat{Z}_{s+1} = y_{\bm{\beta}}(Z_{\leq s})$
based on previous latent states. Thus, the conditional probability
density is equivalent to the density of the prediction residual:
\begin{equation}
  \mathrm{P}(Z_{s+1}\mid Z_{\leq s})
  = \mathrm{P}(\varepsilon_{s+1}\mid Z_{\leq s}),
  \label{eq:residual}
\end{equation}
where $\varepsilon_{s+1}=Z_{s+1} - \hat{Z}_{s+1}$ denotes the residual at time
$s+1$.

The distributions of residuals are usually unknown in practice. In
this paper, we assume they are Gaussian. This assumption could be
inappropriate if the prediction is with the raw RV or when the
transformation $f_{\bm{\theta}}$ is as simple as a Box-Cox
transformation \citep{proietti2013does}. Nevertheless, we find
empirically that a Gaussian distribution is a good approximate for the
residuals after the normalizing flow is optimized, which could be due
to the larger capacity of the normalizing-flow neural network. In
addition, we assume the residuals to be homogeneous across time (i.e.,
$\varepsilon_s$ follows the same Gaussian distribution for every
timestep $s$). Using a more advanced time-series model would improve
the homogeneity of the residuals, which is left to future work.

Therefore, the residuals $\varepsilon_s\sim N(\mu, v)$ ($s=1,
2,\cdots,t$), where $\mu$ and $v$ are the mean and variance of the
Gaussian distribution. $\mu$ is set to zero\footnote{ One may
alternatively relax the prior of $\mu=0$ and set $\mu$ to be empirical
mean instead; however, we did not observe improvements by doing so, as
the empirical mean is usually close to zero.} and $v$ is estimated as
the empirical variance: $v = \left(\sum_{s=1}^t (\varepsilon_s -
\mu)^2 \right) / (t-1)$. Notice that $v$ involves the whole period
$1,2,\cdots,t$, but is used for calculating
$\mathrm{P}(\varepsilon_{s+1}\mid Z_{\leq s})$ for $s<t$, which means
the calculation involves ``further information'' at timestep $s<t$.
Nevertheless, it does not invalidate our approach because the
calculation of $\mathrm{P}(\varepsilon_{s+1}\mid Z_{\leq s})$ is only
required in parameter estimation; that is, making out-of-sample
prediction on $Z_{t+1}, Z_{t+2}, \cdots$ does not involve the
calculation of $\mathrm{P}(\varepsilon_{s+1}\mid Z_{\leq s})$.

\section{Experiment}

\subsection{Data}
\label{sec:data}

The proposed method is evaluated on a dataset of 100 major stocks
listed on the New York Stock Exchange. The dataset covers a continuous
time period of 480 trading days from December 1st, 2015, to October
25th, 2017. The daily RV values were calculated according to Formula
(\ref{eq:rv}) for every stock. Thus, 100 RV time series of 480
timesteps were acquired. In computing the daily RV, we followed
\citet{corsi2009har} and did not include the overnight price
variation. That is, we did not include $p_{t+1,1} - p_{t, n_t}$ in the
summation of Formula (\ref{eq:rv}), where $n_t$ denotes the
number of high-frequency prices within day $t$ as mentioned in Section
\ref{sec:problem}.

In this paper, the RV of a day is calculated with high-frequency
prices that are recorded every five minutes. That is, $p_{t,i+1}$ was
taken five minutes later than $p_{t,i}$. The value of high-frequency
price is calculated as the {\em mid-price} \citep{stoikov2018micro} of
the order book, which is an average of the highest bid price and the
lowest ask price weighted by their volumes, as follows:
\begin{align}
  p_{t,i} &= \alpha_{t,i} p_{t,i}^\text{bid} + (1 - \alpha_{t,i}) p_{t,i}^\text{ask}, \\
  \alpha_{t,i} &= \frac{\text{Vol}(p_{t,i}^\text{bid})}
                       {\text{Vol}(p_{t,i}^\text{ask}) + \text{Vol}(p_{t,i}^\text{bid})},
\end{align}
where $p_{t,i}^\text{bid}$ and $p_{t,i}^\text{ask}$ denote the highest
bid and the lowest ask prices ($p_{t,i}^\text{bid} <
p_{t,i}^\text{ask}$), respectively, at the end of the $i$-th
five-minute interval; $\text{Vol}(\cdot)$ denotes the volume (i.e.,
number of shares) of bid or ask quotes at a certain price. Every RV
time series was z-score standardized (i.e., $x_t \leftarrow (x_t -
\mu) / \sqrt{v}$), where $\mu$ and $v$ are the empirical mean and
variance, respectively, of the times series.

\subsection{Settings}

The RV time series were split into training, validation, and test
sets chronologically, at a ratio of 300, 60, and 120 days.

We conducted the experiments in a univariate time-series setting by
viewing all 100 RV time series as independent samples drawn from the
same random process $X_1,X_2,\cdots$. Thus, this accommodates the
notion of taking an expectation over IID samples in Formula
(\ref{eq:em}). The normalizing flow $f_{\bm{\theta}}$ was optimized on
the training set (i.e., the first 300 days of the RV history of all
stocks). Using an Adam \citep{kingma2014adam} optimizer, we conducted
200 iterations of parameter updates and evaluated the model on the
validation set at every five training iterations, thus producing 40
different snapshots of parameters. The snapshot that achieved the
highest log-likelihood in Formula (\ref{eq:decompose}) was regarded as
the best. In the following, we report the performance of this best
snapshot on the test set.

For the neural network $\text{NN}_{\bm{\theta}}$ used in the NODE (see
Formula (\ref{eq:node})), we used a simple {\em multi-layer
perceptron} with two hidden layers. Each hidden layer had four hidden
units. Formally,
\begin{equation}
  \text{NN}_{\bm{\theta}}(x)=W_3 \sigma(W_2 \sigma(W_1 x + b_1) + b_2) + b_3,
\end{equation}
where $W_1\in\mathbb{R}^{4\times 1}$, $W_2\in\mathbb{R}^{4\times 4}$,
$W_3\in\mathbb{R}^{1\times 4}$, $b_1\in\mathbb{R}^4$,
$b_2\in\mathbb{R}^4$, and $b_3\in\mathbb{R}^1$ are parameters. Thus,
$\bm{\theta}=\{W_1,W_2,W_3,b_1,b_2,b_3\}$ and the transformation has
33 parameters in total. $\sigma$ is an elementwise nonlinear
activation function, and we set it to the {\em swish} function
\citep{ramachandran2017searching} as follows:
\begin{equation}
  \sigma(x) = \frac{x}{1 + \exp(-x)}.
\end{equation} Compared with other popular choices, such as the
sigmoid or the tanh, the swish function has two advantages as follows.
First, it has an unbounded range and thus learns a more natural
transition to transforming large RV values. Second, the swish function
is asymmetric around zero, which facilitates modeling the strong
skewness in RV data.

\subsection{Baselines}
\label{sec:baseline}

We considered different baseliens by varying the transformation
$f_{\bm{\theta}}$ and the distribution for the residuals
$\varepsilon_s$ (see Formula (\ref{eq:har})).

For alternative transformations, we considered the following:
\begin{description}[noitemsep]
  \item[Identity] transformation: $f_{\bm{\theta}}(x)=x$, which
    is equivalent to {\em not} applying a transformation.
  \item[Wallace's] transformation \citep{wallace1959bounds}:
    \begin{equation}
      f_{\bm{\theta}}(x)= \pm \frac{8d+1}{8d+3}\sqrt{d\log(1+|x|^2/d)},
    \end{equation}
    where $\bm{\theta}=\{d\}$.
  \item[Yeo-Johnson] transformation \citep{yeo2000new}: a generalization
    of the Box-Cox transformation \citep{box1964analysis} to allow
    non-positive inputs, defined as follows:
    \begin{equation}
      f_{\bm{\theta}}(x) = \left\{\begin{array}{ll}
      ((x+1)^\lambda - 1) / \lambda  & \text{if}~\lambda\neq 0, x\geq 0, \\
      \log(x+1) & \text{if}~\lambda=0, x\geq 0, \\
      -((-x+1)^{(2-\lambda)} - 1) / (2 - \lambda) & \text{if}~\lambda\neq 2, x<0, \\
        -\log(-x+1) & \text{if}~\lambda=2, x<0,
      \end{array}\right.
    \end{equation}
    where $\bm{\theta} = \{\lambda\}$.
  \item[tanh(k)] transformation \citep{snelson2003warped}:
    a simplest neural network with a single hidden layer, defined as follows:
    \begin{equation}
      f_{\bm{\theta}}(x) = \sum_{i=1}^k u_i \tanh(v_i x+b_i),
    \end{equation}
    where $\bm{\theta}=\bigcup_{i=1}^k \{u_i, v_i, b_i\}$.
    For $f_{\bm{\theta}}$ to be invertible,
    $u_i, v_i$ ($i=1,\cdots,k$) are restricted to be positive.
\end{description}

As for the distribution of the residuals, we tested the generalized
Student's $t$ distribution (in addition to the Gaussian by default),
which is defined by the following probability density function with
mean $\mu$, variance $v$, and $d$ degrees of freedom:
\begin{equation}
  \mathrm{P}(x\mid \mu, v, d) =
  \frac{\Gamma((d+1)/2)}{\sqrt{\pi(d-2) v} \Gamma(d/2)}
  \left(1 + \frac{(x-\mu)^2}{(d-2) v}\right)^{-(d+1)/2}.
\end{equation}
The density function of a Student's $t$ distribution decays much
slower when $x$ goes to infinity; therefore, it is commonly used for
modeling fat-tailed phenomena.

\section{Results}
\label{sec:result}

\subsection{Prediction Accuracy}
\label{sec:accuracy}

Table \ref{tbl:score} summarizes the overall results over the dataset
of 100 stocks acquired with different transformations. Each row
represents a transformation. Our transformation via normalizing flow
is shown at the bottom of the table. The third and fourth columns show
the scores concerning HAR prediction accuracy. RMSE was used as
the primary evaluation metric to follow the previous works
\citep{corsi2009har}. A smaller RMSE implies a higher precision.

The third column of Table \ref{tbl:score} reports the average RMSE
(mean value) over the 100 stocks. One-tailed $t$-tests were conducted
to assess the statistical significance of our method (NODE with
$\tau=0.25$) compared with the other transformations, assuming the
null hypothesis that the mean improvement in RMSE is equal to zero.
Asterisks (*) in the third column indicate statistical significance: *
($p < 0.05$), ** ($p < 0.01$), *** ($p < 0.001$). The fourth column
shows the proportion of stocks on which a method achieved the lowest
RMSE, or ``percentage of the best.'' The best scores are indicated in
bold, and the second-best scores are underlined. For a robustness
test, we also examined NODE with $\tau=5.00$ which has excessive
complexity as a transformation; we omitted the $p$-value and the
``percentage of the best'' value as it serves merely as a robustness
test.

\begin{table}[tbp]
  \centering
  \caption{HAR prediction performance with different transformations
on the dataset of 100 stocks.}
  \label{tbl:score}
  \begin{tabular*}{\linewidth}{l @{\extracolsep{\fill}} ccccc}
    \toprule
    Transformation &
    \multirow{2}{*}{\makecell[c]{Residual distribution\\(presumed)}} &
    \multicolumn{2}{c}{Out-of-sample prediction} &
    \multicolumn{2}{c}{In-sample residual Gaussianity} \\
    \cmidrule(lr){3-4} \cmidrule(lr){5-6}
    & & RMSE & Percentage of best & $R^2$ & Skewness \\
    \midrule
    Identity & Gaussian & 0.5695{***}  & 10\% & 93.22 & 1.788 \\
    \midrule Wallace's & Gaussian & 0.5660{**} & \underline{19\%} & 98.12 & 0.7010 \\
    Yeo-Johnson & Gaussian & \underline{0.5627}{*} & 2\% & 99.09 & 0.2752 \\
    tanh(1) & Gaussian & 0.5642{**} & 6 \% & 99.31 & 0.2059 \\
    tanh(5) & Gaussian & 0.5636{**} & 6 \% & 99.11 & 0.2141 \\
    tanh(10) & Gaussian & 0.5638{**} & 11 \% & 99.20 & 0.1693 \\
    \midrule \multicolumn{3}{l}{Our approach} \\
    NODE($\tau$=5.00) & Gaussian & 0.5630 & - & 99.14 & 0.2141 \\
    NODE($\tau$=0.25) & Gaussian & \textbf{0.5620} & \textbf{46\%} & 98.81 & 0.3928 \\
    \bottomrule
  \end{tabular*}
\end{table}

Our approach (bottom row) achieved the lowest RMSE score at 0.5620, a
clear improvement of 0.0075 compared with the identity transformation
that scored 0.5695. The second-best at 0.5627 was the Yeo-Johnson
transformation, which is a common way of preprocessing non-Gaussian
data in practice; nevertheless, our method still outperformed the
Yeo-Johnson transformation. Compared with the baseline
transformations, the improvements of NODE in RMSE are statistically
significant at the 5\% (Yeo-Johnson), 1\% (Wallace's and $\tanh$
transformations), and 0.1\% (Identity) levels. With respect to the
``percentage of the best,'' the advantage of NODE is more evident.
NODE achieved the lowest RMSE on 46 of the 100 stocks; in comparison,
the Yeo-Johnson transformation won on only two stocks.

The significant margin between NODE and Yeo-Johnson on the
``percentage of the best'' is interesting, in contrast to their
relatively close performance in average RMSE. This indicates that on
many stocks, NODE and Yeo-Johnson transformed realized volatilities
into close distributions, but NODE performed slightly better in most
cases. While power transformations like Yeo-Johnson are careful
choices made by practitioners after decades, NODE was learned from
data. However, NODE ``discovered'' the power transformations were the
best choices and further improved over them.

The transformations denoted by tanh($k$) ($k=1,5,10$) are the simplest
feed-forward neural networks, and $k$ represents the number of hidden
units. When $k$ was increased from 1 to 5, the average RMSE was
improved from 0.5642 to 0.5636. However, when $k$ was further
increased to 10, the score degraded to 0.5638.

The degradation at large $k$ values might be caused by the increased
risk of overfitting to the training set. The tanh transformations
with a large $k$ or NODE with a large $\tau$ had a stronger
approximation capability, but they overfitted to the training set.
Evidence of this is that on 29 stocks, simpler transformations,
including Identity or Wallace's, had the best RMSE. On these stocks,
the data distribution of the test set shows inconsistency with the
training set, which is common in a financial market that is a
non-stationary system. Such overfitting is also seen for NODE when
$\tau$ was increased from 0.25 to 5.00 when RMSE increased from 0.5620
to 0.5630.

\begin{figure*}[t]
  \centering
  \includegraphics[width=0.35\linewidth]{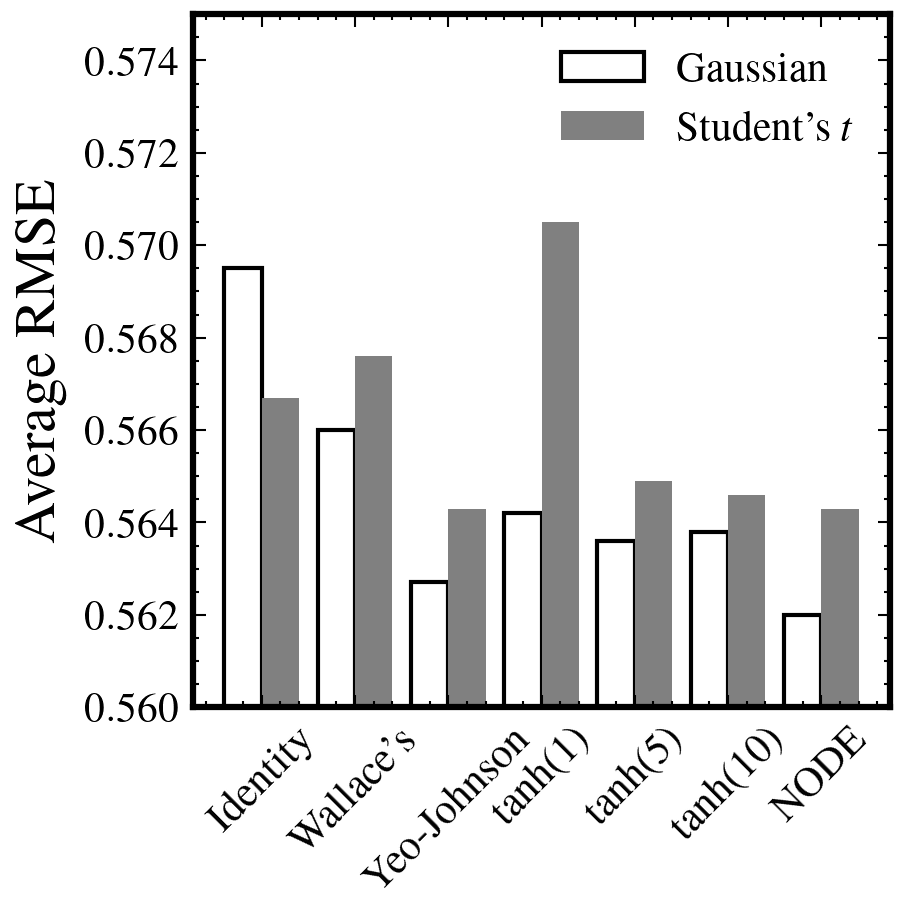}
  \caption{A comparison between two distributional
  assumptions (Gaussian or Student's $t$) on the residuals
  $\varepsilon_s$ in Formula (\ref{eq:har})
  under different analytical/neural transformations (horizontal axis),
  regarding the average RMSE over 100 stocks.}
  \label{fig:rmse-resdist}

  \vskip 1em
  \includegraphics[width=0.95\linewidth]{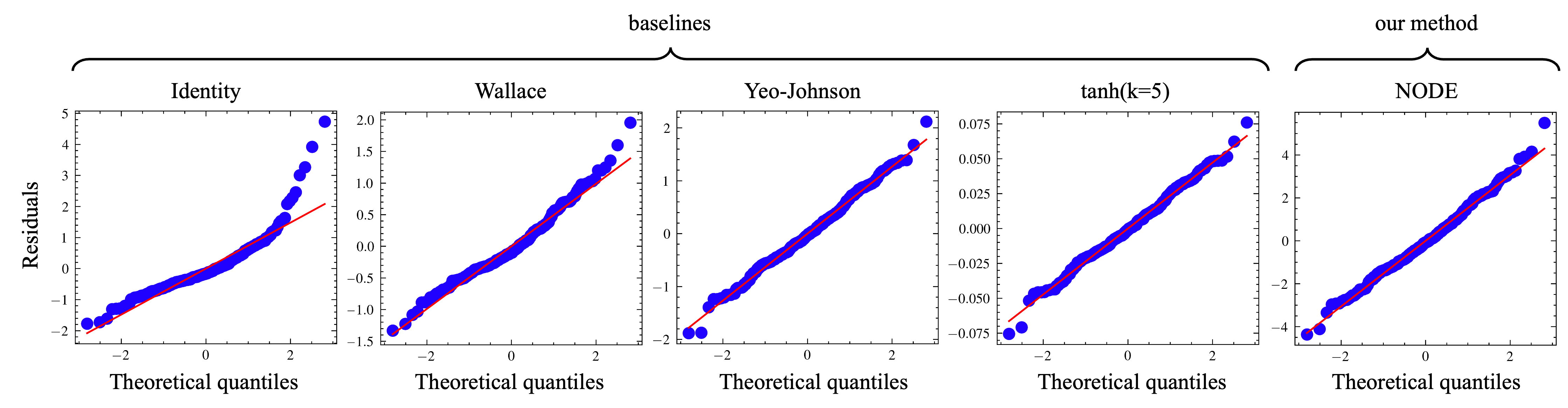}
  \caption{Comparison between the percentiles of residuals (vertical
  axes) and the standard normal distribution (horizontal axes). Each
  plot represents a transformation. The straight red lines
  represent linear fits to the data points within each plot.
  } \label{fig:residual}
\end{figure*}

As an ablation analysis, we evaluated the robustness of the
presumption made in Section \ref{sec:residual} that the residuals of
HAR (i.e., $\varepsilon_s$ in Formula (\ref{eq:har})) follow a
Gaussian distribution. Instead of assuming that $\varepsilon_s\in
N(\mu, v)$, the ablated version assumes a generalized Student's $t$
distribution with an additional parameter $d$ denoting the degrees of
freedom. Note that $d\to\infty$ recovers a Gaussian distribution.

Figure \ref{fig:rmse-resdist} shows the average RMSE scores (vertical
axis) when the residual distribution was assumed to be either Gaussian
(white bars) or generalized Student's $t$ (grey bars) grouped by the
transformation $f_{\bm{\theta}}$. When the residuals are
assumed to follow a generalized Student's $t$ distribution, the
probability density in Equation (\ref{eq:residual}) is determined
using the PDF of the Student's $t$ distribution, instead of the
Gaussian PDF. Consequently, the transformation parameters
$\bm{\theta}$ that are estimated yield an RMSE score that differs from
what would be expected under a Gaussian residual assumption.

In Figure \ref{fig:rmse-resdist}, the white bars are the values in
Table \ref{tbl:score} in the third column. Except for the
Identity transformation, we observed increased RMSE scores when the
presumed distribution was a Student's $t$, which means a decrease in
prediction accuracy. The largest increase in RMSE is observed with the
tanh(1) transformation, implying tanh(1) to be sensitive to the
distributional assumption. In contrast, NODE still achieved the
smallest RMSE score at 0.5643, the same as with a Yeo-Johnson
transformation. The results in Figure \ref{fig:rmse-resdist} suggest
the robustness of our approach even under a generalized Student's $t$
distribution assumption for residuals that is not optimal.

\subsection{Qualitative Comparison}

\begin{figure*}
  \centering
  \includegraphics[width=0.4\linewidth]{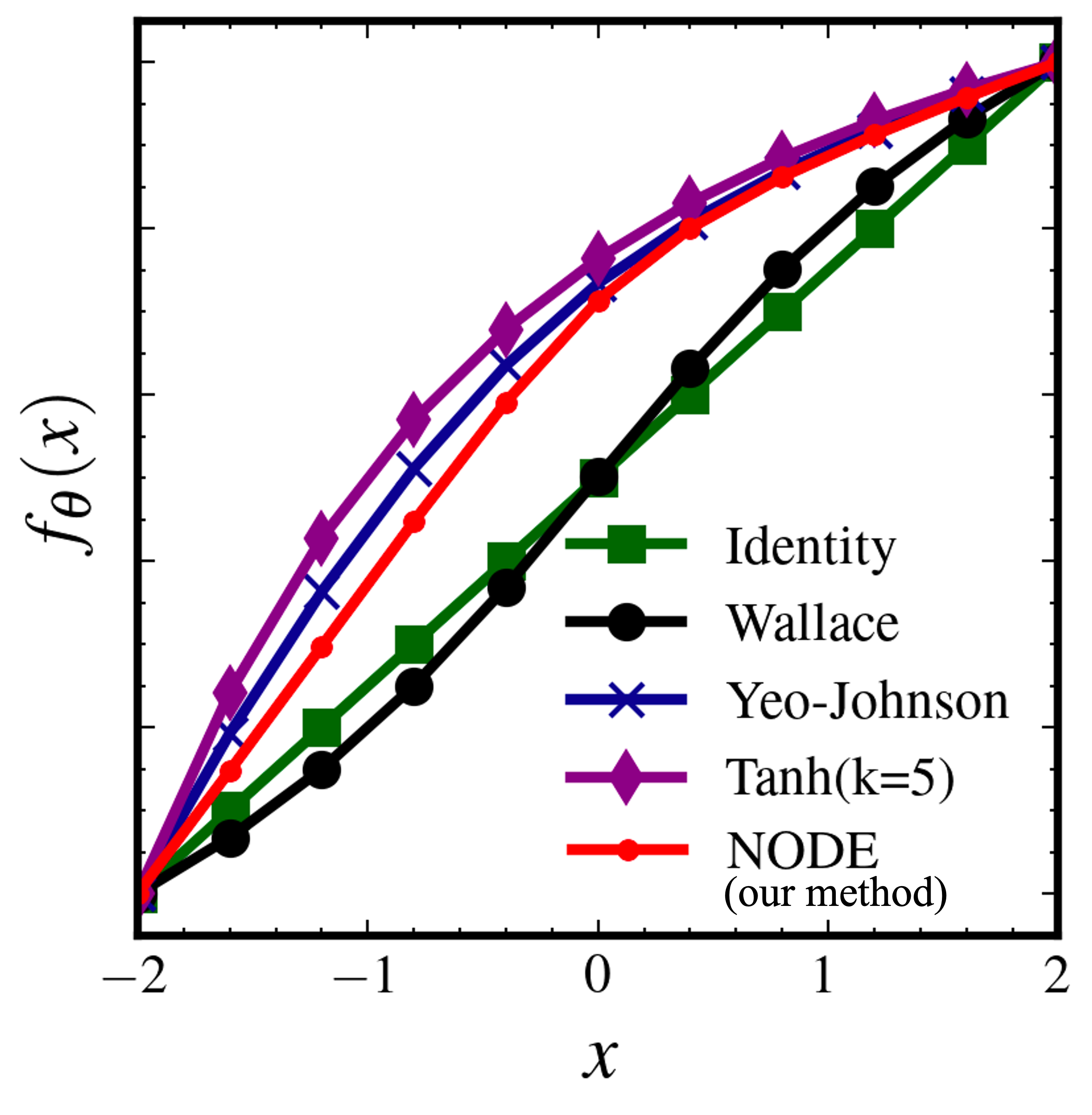}
  \caption{The graph of $f_{\bm{\theta}}(x)$ vs. $x$
    within $x\in[-2, 2]$.
    Each plot represents a transformation. For every transformation,
    $f_{\bm{\theta}}(x)$ was standardized so that
    $f_{\bm{\theta}}(-2)=-2$ and $f_{\bm{\theta}}(2)=2$. The range
    $[-2, 2]$ covered most RV values that were
    z-score standardized in preprocessing.
  }
  \label{fig:fx}
\end{figure*}

Figure \ref{fig:residual} provides a comparison between the
distribution of residuals $\varepsilon_s$ defined in Formula
(\ref{eq:har}) (vertical axes) and the standard normal
distribution (horizontal axes) for the stock ``BABA'' (i.e., the Q-Q
plot). Each plot represents a different transformation listed in
Section \ref{sec:baseline}. In each plot, a point represents the
residual at a timestep; the vertical axis shows the residual values,
and the horizontal axis shows the theoretical percentiles if the
residuals follow a standard normal distribution. The straight red line
is a linear fit of the points, and a better fit implies a higher
degree to which the residuals follow a Gaussian distribution.

As seen in the first plot of Figure \ref{fig:residual} (Identity
transformation), the residuals of the raw RV deviate from the linear
fit at large positive quantiles. This shows how HAR fails to fully
capture the skewness in RV time series. In contrast, with a
transformation, the Gaussianity of the residuals was largely improved.
This improvement is visible for all the transformations shown in
Figure \ref{fig:residual}. Among the transformations tested, the
Yeo-Johnson, the tanh(5), and NODE produced impressive Gaussianity, and
they are indistinguishable within a large range.

The Gaussianity of the residuals is further quantified using two
metrics, as shown in the two right-hand columns in Table
\ref{tbl:score}. The first metric is the $R^2$ score of the linear fit in
the Q-Q plots of Figure \ref{fig:residual}, and the other is the
skewness (i.e., the third moment) of the residuals. The two right-hand
columns present the average $R^2$ and the average skewness over the
100 stocks.

Consistent with the observation of Figure \ref{fig:residual}, without
a transformation (first row), HAR produced residuals with a low $R^2$
at 93.22\% and a high skewness at 1.7883, implying poor Gaussianity.
The Gaussianity was greatly improved by using a transformation. $R^2$
was improved to 99.14\%, and skewness was reduced to 0.2141 with a
NODE having $\tau=5.00$ (second to last row).

When $R^2$ exceeds 99\%, an improvement in $R^2$ or a reduction in
skewness did not translate into an improvement in out-of-sample RMSE.
This corresponds with our previous conjecture on overfitting.
Nevertheless, even with the same level of $R^2$ and skewness, NODE
with $\tau=5.00$ still outperformed the tanh(5) transformation in
out-of-sample RMSE.

Figure \ref{fig:fx} provides a visualized comparison between the
transformations on the real line. The horizontal axis represents the
input $x\in\mathbb{R}$ to the transformation $f_{\bm{\theta}}$, and
the vertical axis shows the output (i.e., $f_{\bm{\theta}}(x)$). Note
that the realized volatilities were z-score standardized in
preprocessing, which produced negative $x$. For visualization, we also
linearly rescaled the transformations so that $f_{\bm{\theta}}(-2) =
-2$ and $f_{\bm{\theta}}(2) = 2$. As HAR is a linear model with a bias
term, the estimation and prediction with HAR are invariant under
linear rescaling to the transformation.

Each curve in Figure \ref{fig:fx} represents a transformation. The
straight green line represents the identity function, and the dotted
red plot represents the NODE transformation. The NODE, Yeo-Johnson (in
blue), and tanh transformations (in purple) all showed a concave
shape: their slopes gradually decrease as $x$ increases. This
corresponds with our expectation that larger realized volatilities are
calibrated more than smaller realized volatilities to eliminate
skewness.

At large $x$, NODE showed almost perfect consistency with the
Yeo-Johnson transformation. However, a discrepancy is seen at small
$x$ where NODE grows linearly, similar to the Identity function. In
other words, NODE viewed it ``unnecessary'' to apply nonlinear
calibration to small realized volatilities. With the Yeo-Johnson or
tanh transformations, such local linear growth is not possible, as
power and tanh transformations are defined globally over the real
line.

\begin{figure}
  \centering
  \includegraphics[width=0.45\linewidth]{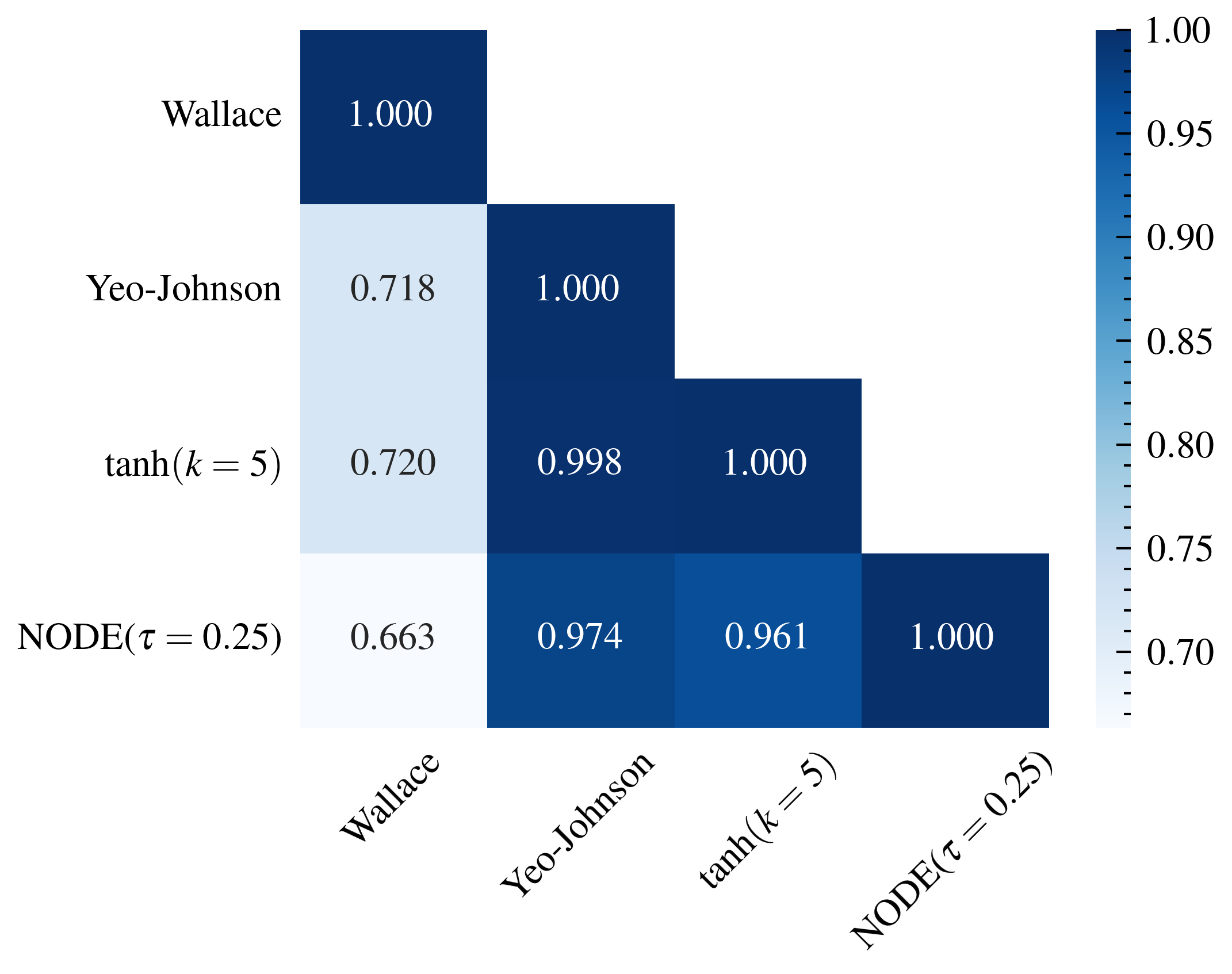}
  \caption{Pearson's correlation coefficients between the
improvements in RMSE for various transformations over the Identity
transformation, calculated from the results of the 100 stocks. }

  \label{fig:corr}
\end{figure}

Figure \ref{fig:corr} presents the Pearson's correlation coefficients
between the transformations on the RMSE improvements over the
Identity transformation, measured for the 100 stocks. The RMSE values
were obtained through out-of-sample evaluations, and the mean value
for each transformation has been reported in Table \ref{tbl:score},
in the third column.

These correlation results align closely with the graph of
$f_\theta(x)$ in Figure \ref{fig:fx}. Notably, the Yeo-Johnson and the
$\tanh(k=5)$ transformations exhibit a correlation coefficient of
0.998. This is unsurprising as they appear nearly identical in shape
in Figure \ref{fig:fx} (the purple and blue plots). Furthermore, NODE
also demonstrates a high correlation with the Yeo-Johnson
transformation, which validates our conjecture in Section
\ref{sec:accuracy} that NODE has ``discovered'' (while also
outperformed) power transformations from the data.

\section{Conclusion}

We proposed a new method to enhance the prediction of realized
voltility by co-training a simple linear prediction model with a
nonlinear transformation. In contrast to previous methods that
estimate the transformation before the prediction model and use
separate objective functions at the two steps, we propose to co-train
the two parts following a unified maximum-likelihood objective
function. Additionly, we introduced a method based on the
expectation-maximization algorithm to jointly estimate the parameters
for both parts.

For the nonlinear transformation, we incorporated normalizing flows
which represent the state-of-the-art in neural distributional
transformations. We demonstrated how the proposed co-training
procedure can utilize complex transformations, a task challenging for
prior methods restricted to simple analytical functions.

On a dataset of the high-frequency price history of 100 stocks for two
years (2015-2017), the proposed method significantly outperformed
predictions with the raw time series in average RMSE. Compared with
analytical and neural-network baselines, our method achieved the best
RMSE on 46 of the 100 stocks, suggesting its effectiveness and
robustness.

\begin{acks}
  This work was supported by \grantsponsor{JSPS}{JSPS
  KAKENHI}{} Grant Numbers \grantnum{JSPS}{JP20K20492} and
  \grantnum{JSPS}{JP21H03493}.
\end{acks}

\bibliographystyle{ACM-Reference-Format}
\bibliography{main}

\end{document}